\documentclass[twocolumn,aps,amsmath,amssymb,showpacs,showkeys,floatfix,prl,superscriptaddress]{revtex4}

\usepackage{bbm}
\usepackage[dvips]{graphicx}
\usepackage{color}
\usepackage{amsmath,amssymb,latexsym}

\newcommand{\eps}{\mathcal{E}}

\newcommand{\rom}[1]{\uppercase\expandafter{\romannumeral #1\relax}}
\DeclareMathOperator{\starup}{^{\text{*}}}

\renewcommand{\ol}[1]{{\langle #1\rangle}}

\begin{document}

\title{Scaling and Regeneration of Self-Organized 
Patterns}

\author{Steffen Werner}
\affiliation{Max-Planck-Institute for the Physics of Complex Systems, Nšthnitzer Stra§e 38, 01187 Dresden, Germany}
\author{Tom St{\"u}ckemann}
\affiliation{Max-Planck-Institute of Molecular Cell Biology and Genetics, Pfotenhauerstr. 108, 01307 Dresden, Germany }
\author{Manuel Beir\'an Amigo}
\affiliation{Max-Planck-Institute for the Physics of Complex Systems, Nšthnitzer Stra§e 38, 01187 Dresden, Germany}
\affiliation{Universidad Aut\'onoma de Madrid, Ciudad Universitaria de Cantoblanco, 28049 Madrid, Spain}
\author{Jochen C. Rink}
\affiliation{Max-Planck-Institute of Molecular Cell Biology and Genetics, Pfotenhauerstr. 108, 01307 Dresden, Germany }
\author{Frank J\"ulicher}
\affiliation{Max-Planck-Institute for the Physics of Complex Systems, Nšthnitzer Stra§e 38, 01187 Dresden, Germany}
\author{Benjamin M. Friedrich}
\email{benjamin.friedrich@pks.mpg.de}
\affiliation{Max-Planck-Institute for the Physics of Complex Systems, Nšthnitzer Stra§e 38, 01187 Dresden, Germany}

\date{\today}



\begin{abstract}
Biological patterns generated during development and regeneration often scale with organism size. Some organisms, e.g., flatworms, can regenerate a rescaled body plan from tissue fragments of varying sizes. Inspired by these examples, we introduce a generalization of Turing patterns that is self-organized and self-scaling. 
A feedback loop involving diffusing expander molecules regulates the reaction rates of a Turing system, thereby adjusting pattern length scales proportional to system size.
Our model captures essential features of body plan regeneration in flatworms as observed in experiments.
\end{abstract}

\maketitle


Understanding the morphogenesis of a complex multicellular organism from a single fertilized egg
poses a fundamental challenge in biology \cite{wolpert2011principles,gilbert2014developmental}.
The diversity of shapes of living organisms emerges from biological patterning processes that assign cell fates 
depending on the spatial position of cells \cite{wolpert2011principles}. 
Patterning processes are remarkably precise and reproducible, 
despite environmental perturbations and 
the stochastic nature of fundamental cellular processes such as gene expression \cite{abouchar2014fly}. 
Furthermore, the astonishing regeneration capabilities of certain animals, 
including flatworms, polyps, salamanders, and newts, require patterning mechanisms that additionally can cope with highly variable initial conditions
\cite{liu2013reactivating,galliot2010hydra,voss2009ambystoma,eguchi2011regenerative}. 
Both the robust establishment and the scaling of patterns during growth 
are poorly understood.

The fruit fly \textit{Drosophila melanogaster} 
has been an important model system to study biological pattern formation
and body plan scaling \cite{bate1993development, stjohnston1992origin, gregor2005diffusion, umulis2013mechanisms}.
There, specific molecules, called morphogens, are secreted in localized source regions. 
Morphogens establish long-range concentration profiles by the interplay of transport and degradation.
They provide chemical signals away from the source that can regulate patterning and growth 
\cite{othmer1980scale, affolter2007decapentaplegic, wartlick2009morphogen, benzvi2010scaling, wartlick2011dynamics, wartlick2011understanding, benzvi2011expansion, wartlick2014growth}.
Specifically, fly wing development has been extensively studied \cite{affolter2007decapentaplegic, wartlick2009morphogen, wartlick2011dynamics, wartlick2011understanding, benzvi2011expansion, hamaratoglu2011dpp}.
Quantification of morphogen profiles in the developing fly wing 
at different stages of development
revealed that the morphogen concentration profiles scale with the size of the growing tissue, 
maintaining an approximately constant shape \cite{wartlick2011dynamics, wartlick2011understanding, benzvi2011expansion, hamaratoglu2011dpp}.
In a minimal description, the characteristic decay length 
$\lambda=(D/k)^{1/2}$ of these concentration profiles
depends on the effective diffusion coefficient $D$
and the degradation rate $k$ \cite{gregor2005diffusion, wartlick2009morphogen}.
It has been proposed that the scaling of these profiles is achieved
by a dynamic regulation of the morphogen degradation rate
via a chemical signal, called the expander, whose level varies with system size
\cite{othmer1980scale, benzvi2010scaling, wartlick2011dynamics, wartlick2011understanding, benzvi2011expansion}.
Different possible realizations for such mechanisms have been proposed 
\cite{umulis2013mechanisms, othmer1980scale, benzvi2010scaling, wartlick2011dynamics, wartlick2011understanding, benzvi2011expansion, benzvi2011scaling, hunding1988size, ishihara2006turing}.
These mechanisms rely on prepatterned tissues with specified sources or sinks for morphogens or the expander.

Scaling and regeneration of the entire body plan in the flatworm {\it Schmidtea mediterranea} 
challenges scaling mechanisms that rely on prepatterned cues. 
{\it Schmidtea mediterranea} can regenerate the complete animal from minute tissue fragments 
by repatterning the fragment to establish a proportionately scaled body plan \cite{newmark2002not}. 
Furthermore, flatworms grow when fed and literally shrink when starving, 
scaling their body plan proportionally over more than one order of magnitude in length
($\approx 0.5 - 20$ mm for {\it Schmidtea mediterranea}) \cite{newmark2002not}. 
These experimental observations prompt the existence of patterning systems with remarkable self-organizing and self-scaling properties.
Recently, chemical signals have been identified whose perturbations have long-range effects on body plan patterning and regeneration.
In particular, Wnt signaling, 
a pathway with conserved roles for developmental patterning, 
determines head-tail polarity during flatworm regeneration
\cite{gurley2008betacatenin,adell2010gradients,almuedocastillo2012wnt,gurley2010expression}. 
Inspired by these examples of biological pattern formation,
we address in this Letter general requirements for 
the emergence of robust patterns that scale with system size.

\begin{figure}[tbp]
\centering
\includegraphics[width=0.45\textwidth]{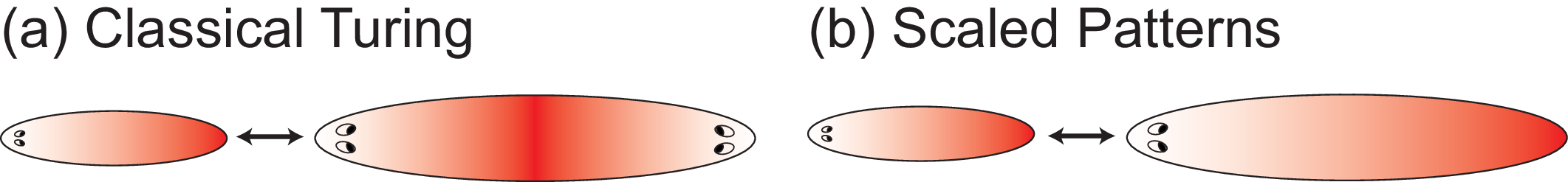}
  \caption{\small 
Classical Turing patterns show more periodic repeats in larger systems 
as a result of fixed intrinsic length scales (a), 
instead of being a scaled-up version of the patterns in small systems (b).} 
  \label{FishFixedLengthScale}
\end{figure}\noindent

The simplest model to spontaneously generate head-tail polarity based on graded concentration profiles of signaling molecules 
is the classical reaction-diffusion system introduced by Turing 
\cite{turing1952chemical,gierer1981generation, meinhardt2009models}.
However, the resulting patterns do not scale naturally as sketched in Fig.~\ref{FishFixedLengthScale}, 
since diffusion coefficients and reaction rates define fixed characteristic length scales.
Here, we extend the Turing model and introduce a self-organized feedback 
mediated by an expander molecule. 
This allows the system to robustly scale concentration profiles and source regions
over several orders of magnitude of system size.
Our model illustrates a general mechanism that could account 
for essential features of pattern scaling and regeneration observed in biological systems.\\

\paragraph{\it Size dependence and multistability of Turing patterns.}
We briefly recall the classical Turing framework to highlight the size dependence of its emergent patterns and to introduce the notation used throughout this Letter. We consider a minimal version of the Turing mechanism, 
which consists of two chemical species (with concentrations $A$ and $B$)
that diffuse with diffusion coefficient $D_A$ and $D_B$ and interact in a one-dimensional domain of size $L$ with reflecting boundary conditions
\begin{eqnarray}
\label{EqReactDiffChoice}
\partial_t A&=&\alpha_A\,P(A,B)-\beta_A\,A+D_A\,\partial_x^2\,A\nonumber\\
\partial_t B&=&\alpha_B\,P(A,B)-\beta_B\,B+D_B\,\partial_x^2\,B\,.
\end{eqnarray}
We specifically consider linear degradation with rates $\beta_A$ and $\beta_B$ and production with rates $\alpha_A$ and $\alpha_B$, and a switch-like Hill-function typical for cooperative and competitive chemical reactions in biological systems:
\begin{equation}
P(A,B)=\frac{A^h}{A^h+B^h}\,.
\label{EqHill}
\end{equation}
Equation (\ref{EqHill}) implies that production is switched on if the activator concentration $A$ exceeds the inhibitor concentration $B$.
The choice of Eqs.~\ref{EqReactDiffChoice} and \ref{EqHill} is conceptually equivalent to Turing's original formulation 
\cite{turing1952chemical}, yet particularly suitable for analytical treatment.
The diffusion coefficients and degradation rates define two characteristic length scales
\begin{equation}\lambda_A=\sqrt{D_A/\beta_A}\;,\quad \lambda_B=\sqrt{D_B/\beta_B}\,.\label{EqLambdas}
\end{equation}
The interplay between these length scales and the system size determines the final patterns, as we show next.

\begin{figure}[tbp]
\centering
\includegraphics[width=0.48\textwidth]{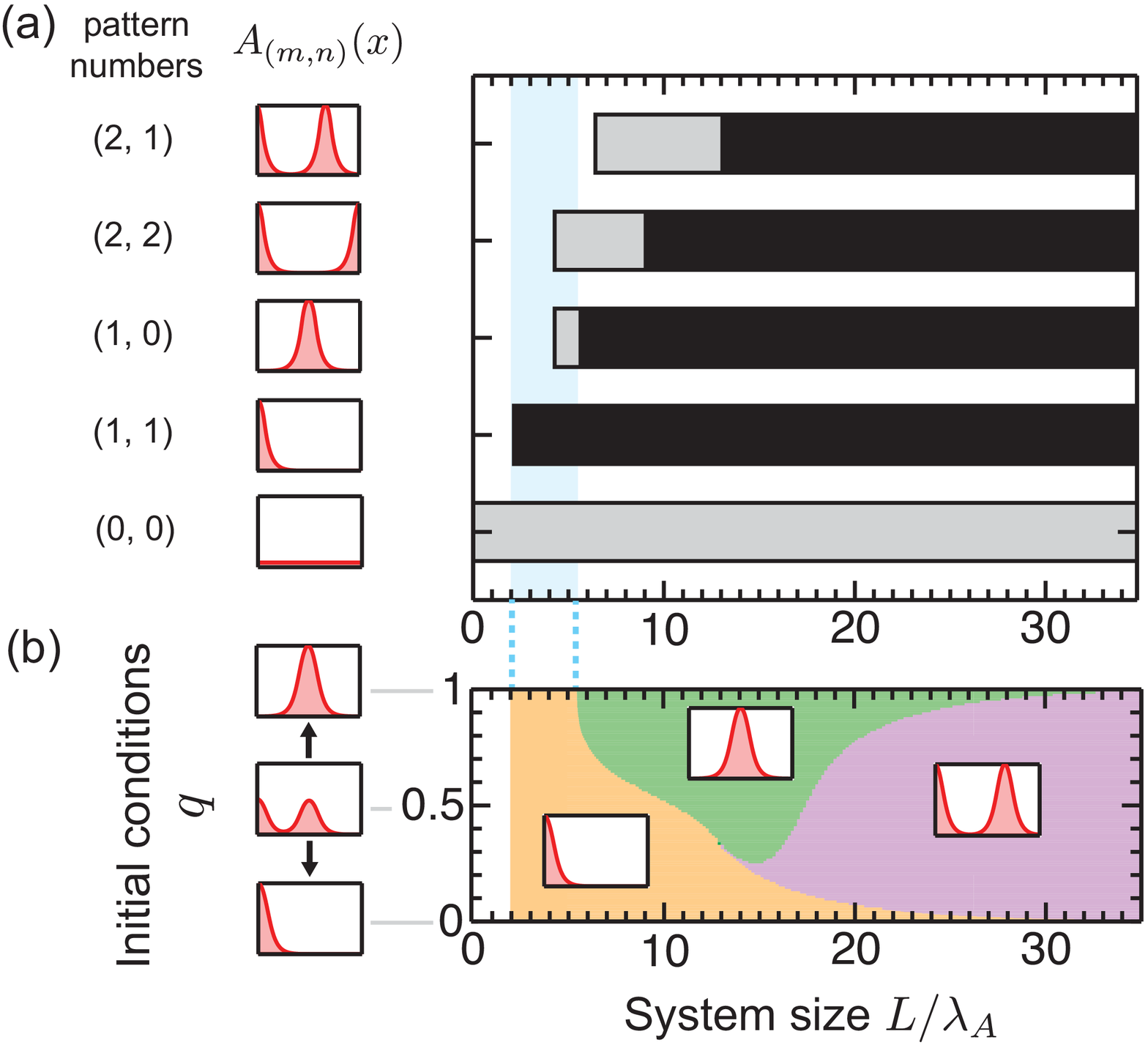}
\caption[]{\small 
Classical Turing patterning implies that 
in larger systems higher-order patterns form.
(a)
Steady-state patterns of Eq.~(\ref{EqReactDiffChoice}) are classified by two pattern numbers $(m,n)$:
$m$ is the total number of contiguous source regions, 
while $n$ is the number of source regions touching the system boundaries. 
Typical profiles of the activator concentration $A_{(m,n)}(x)$ for the $(m,n)$-pattern are shown in red.
Size ranges are shown, where the $(m,n)$-pattern is linearly stable (black),
or exists, but is not stable (gray).
In the blue region, the $(1,1)$-pattern is the only stable pattern.
(b)
Basins of attraction: 
final pattern type at steady state as a function of system size on the horizontal axis and initial conditions on the vertical axis.
Initial conditions linearly interpolate 
between the (1,1)- and (1,0)-pattern, i.e.,
$A(x,t{=}0){=}(1{-}q)A_{(1,1)}(x){+}qA_{(1,0)}(x)$, 
and analogously for $B(x,t{=}0)$.
Parameters: 
$D_B/D_A = 30$, $\alpha_B/\alpha_A = 4$, $\beta_B/\beta_A = 2$,
$h{\rightarrow}\infty$ (a), $h = 5$ (b).
}
  \label{ModeStabPlot}
\end{figure}

Equation (\ref{EqReactDiffChoice}) possesses a unique homogeneous steady state, 
which can become unstable with respect to inhomogeneous perturbations \cite{turing1952chemical,gierer1981generation, meinhardt2009models}.
For $h{\rightarrow}\infty$, corresponding to a binary source switch $P(A,B)=\Theta(A-B)$, 
we can analytically solve for all inhomogeneous steady-state patterns of Eqs.~(\ref{EqReactDiffChoice}) and (\ref{EqHill}).
These are indexed by the number $m$ of contiguous sources,
defined as regions in which $A>B$,
and the number $n$ of source regions touching the system boundaries, see Fig \ref{ModeStabPlot}(a).
In fact, the $(m,n)$-pattern can be constructed as the concatenation of
$2\,m-n$ copies of the (1,1)-pattern, which thus serves as a basic building block.
The $(m,n)$-pattern exists only if $L$ exceeds a critical size that linearly increases with mode number $2m-n$ (gray region).

We numerically find that steady-state patterns become linearly stable only above a second critical size (black region).
In large systems, several stable steady states coexist.
However, in systems of increasing size, 
we observed increasingly smaller basins of attraction of patterns with small mode number,
rendering these patterns unstable with respect to finite-amplitude perturbations,
as exemplified in Fig.~\ref{ModeStabPlot}(b).

The $(1,1)$-pattern is globally stable only in a limited size range, see Fig.~\ref{ModeStabPlot}(a) (blue region).
Next, we show how the introduction of a third reaction species $E$ stabilizes the (1,1)-pattern, irrespective of system size.\\

\noindent

\begin{figure}[h!]
\centering
\includegraphics[width=0.45\textwidth]{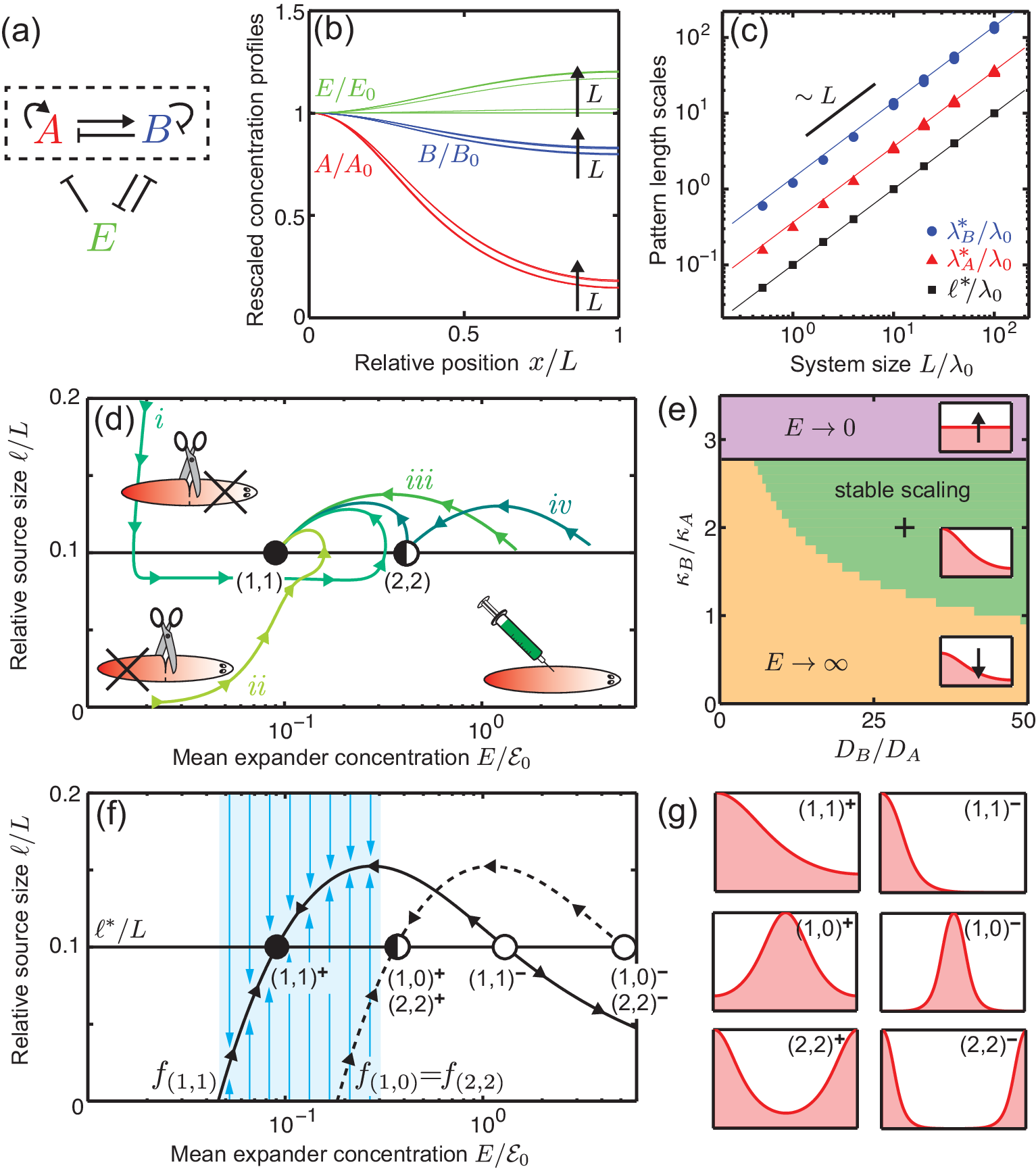}
\caption{\small Scalable pattern formation in a Turing system with expander feedback. 
(a) 
The Turing system and the expander mutually control their degradation rates, resulting in a stable feedback loop. 
(b) 
Scaling corresponds to morphogen profiles that collapse as a function of relative position $x/L$ 
(normalized by respective concentrations $A_0$, $B_0$, $E_0$ at $x=0$).
(c) 
The feedback self-consistently adjusts the length scales $\lambda_A$ and $\lambda_B$ of the morphogen profiles 
and thus the source size $\ell$ with system size
(symbols: numerical results; lines: analytical solution of Eqs.~(\ref{EqReactDiffChoice}) and (\ref{EqEdot}) at steady state for homogeneous expander concentration and $h{\rightarrow}\infty$).
Here,
$\eps_0=(\alpha_A/\kappa_A)^{1/2}$ and
$\lambda_0=[D_A/(\eps_0\kappa_A)]^{1/2}$ denote the characteristic concentration and length scales of the system. 
(d) 
Example trajectories, mimicking amputation experiments (labeled \textit{i},\textit{ii}), 
and uniform, one-time injection of the expander (labeled \textit{iii},\textit{iv}); 
all converge to the same stable fixed point, an appropriately scaled (1,1)-pattern.
(e)
Parameter regions for stable, self-scaling pattern formation (green),
and regions of expander divergence (orange, purple). Parameters of panels (a)-(d) indicated by cross.
(f)-(g)
For adiabatically slow expander dynamics, the system relaxes along the nullclines of the Turing system $f_{(m,n)}$ 
(shown for $h\rightarrow\infty$, $\lambda_E\gg L$). 
As each nullcline intersects the steady-state condition of Eq.~\ref{Eqrelsourceconst} twice, the system possesses two fixed points $(n,m)^+$ and $(n,m)^-$ for each pair $(n,m)$. In the blue region, the (1,1)-pattern is the only stable steady state of the Turing system, compare to Fig.~\ref{ModeStabPlot}, implying that all trajectories must converge to this fixed point.
Parameters: 
$D_B/D_A=30,\quad D_E/D_A=10,\quad \alpha_B/\alpha_A=4,\quad\alpha_E/\alpha_A=0.4,\quad\kappa_B/\kappa_A=2,\quad\kappa_E/\kappa_A=2,\quad 
h=5$, $L/\lambda_0=10$, unless indicated otherwise.
}
\label{SourceGradientScaling}
\end{figure}

\paragraph{\it Pattern scaling by gradient scaling.}
We present a specific example for a general class of minimal feedback mechanisms that yield pattern scaling by adjusting the intrinsic pattern length scales $\lambda_A$ and $\lambda_B$. 
A third molecular species $E$, termed the expander, is produced homogeneously, diffuses, and is subject to degradation
\begin{equation} \partial_t E=\alpha_E-\kappa_{E}\,B\,E+D_E\,\partial_x^2\,E\,.\label{EqEdot}
\end{equation}
The Turing system controls the degradation rate of the expander via the inhibitor $B$. In turn, the expander shall feedback on the Turing system, see Fig.~\ref{SourceGradientScaling}(a). We choose a regulation of the degradation rates by the expander (with $\kappa_{A}$, $\kappa_{B}>0$)
\begin{equation} 
\beta_A=\kappa_{A}\, E\;,\quad  \beta_B=\kappa_{B}\, E\,.
\label{EqbeE}
\end{equation}
We define the relative source size 
$\ell/L=\ol{P}$ and expander-dependent pattern length scales 
$\lambda_A=(D_A/\ol{\kappa_AE})^{1/2}$ and
$\lambda_B=(D_B/\ol{\kappa_BE})^{1/2}$, 
analogous to Eq.~(\ref{EqLambdas}).
Here, the brackets denote spatial averages over the system.

We numerically find that the source size of steady-state patterns scales with system size over several orders of magnitude, see Figs.~\ref{SourceGradientScaling}(b)-(c). 
Concomitantly, we obtain a scaling of the effective Turing length scales $\lambda^*_A\propto L$ and $\lambda^*_B\propto L$, 
where the asterisk denotes steady state.

\noindent

We can challenge pattern scaling by perturbations that mimic experiments such as amputations, 
see Fig.~\ref{SourceGradientScaling}(d).
Two example trajectories,
corresponding to head and tail fragments, respectively, 
converge to an appropriately rescaled (1,1)-pattern,
after a transient overshoot of the source size.
Two additional trajectories, simulating uniform injection of the expander, likewise converge to this fixed point.
One trajectory [labeled \textit{iv} in Fig.~\ref{SourceGradientScaling}(d)] 
is characterized by the transient formation of a second source.

We observe pattern scaling for a vast parameter range,
provided (i) inhibitor diffusion is sufficiently fast (a necessary
condition for pattern formation in any Turing system)
and (ii) the expander feedback strength falls into an intermediate range,
see Fig.~\ref{SourceGradientScaling}(e).

Next, we provide insight into how and why scaling works.
First, we identify steady states, each of which scales with system size. 
For the simple case of adiabatically slow expander dynamics, 
we then show that the (1,1)-pattern is a stable steady state.

The extended Turing system with expander feedback generates steady states, 
for which the relative source size $\ell^*/L$ is independent of system size $L$. 
This can be shown from Eqs.~(\ref{EqReactDiffChoice}) 
and (\ref{EqEdot}) at steady state. By spatial averaging, 
we obtain $0=\alpha_B \ol{P^*} - k_B \ol{B^* E^*}$ and 
$0=\alpha_E - k_E\ol{B^*E^*}$ and hence
\begin{equation}
\frac{\ell\starup}{L}=\frac{\alpha_E\,\kappa_B}{\alpha_B\,\kappa_E}\,.
\label{Eqrelsourceconst}
\end{equation}
In addition, also the pattern length scales $\lambda^*_A$ and
$\lambda^*_B$ scale with high precision with system size. 
In the limit of large expander range [$\lambda_E=(D_E/\ol{\kappa_E B})^{1/2}\gg L$], 
for which the concentration profile of $E$ is approximately homogeneous, scaling becomes exact.
For simplicity, we consider a binary source switch ($h{\rightarrow}\infty$). 
If the expander level was imposed as constant $E=E_0$, 
the Turing system would reach one of the $(m,n)$-patterns discussed above 
in the absence of expander feedback,
with pattern length scales $\lambda_A(E_0)$ and $\lambda_B(E_0)$.
The relative source size $f_{(m,n)}=l/L$ of such a pattern depends on $E_0$
only via the dimensionless ratios $\lambda_A(E_0)/L$ and $\lambda_B(E_0)/L$.
Hence, $f_{(m,n)}=f_{(m,n)}(L^2 E_0)$ is a function of $L^2E_0$.
This shows that changing $E_0$ has analogous effects on the relative source size as changing $L$ in the classical Turing system. 
The same argument also implies that a $(m,n)$-pattern can only exist above a critical value of $E_0$, 
corresponding to the minimum system size for the existence of patterns in Fig.~\ref{ModeStabPlot}(a). 
Below this critical value, $f_{(m,n)}$ is zero. 
Above this value, $f_{(m,n)}$ displays a nonmonotonic dependence on $E_0$, 
which results from opposing effects of the 
pattern length scales of the activator and the inhibitor on the source size $\ell$,
see Fig.~\ref{SourceGradientScaling}(f).
The intersections of the curves $f_{(m,n)}$ with the
constant value $\ell^*/L$ given by Eq.~(\ref{Eqrelsourceconst}) define the steady states of the full system with
expander feedback.
For each pattern type $(m,n)$, 
we find two steady-state patterns, denoted $(m,n)^+$ and $(m,n)^-$, with respective expander levels $E^{+}_{(m,n)}<E^{-}_{(m,n)}$, 
see the black and white circles in Fig.~\ref{SourceGradientScaling}(f).

The fact that $f_{(m,n)}(L^2\,E^*)=\ell^*/L$ is independent of system size $L$ by Eq.~(\ref{Eqrelsourceconst}), implies that also $L^2E^*$ is independent of $L$ for each steady state. 
We conclude $E^*\propto L^{-2}$ and thus 
$\lambda_A(E^*)\propto L$, $\lambda_B(E^*)\propto L$, 
consistent with our numerical results in Fig.~\ref{SourceGradientScaling}(c).

We now discuss the stability of the (1,1)-pattern in the
simple limit of adiabatically slow expander feedback. In this limit,
the source size first relaxes to $\ell/L=f_{(m,n)}(L^2\,E)$ for some
$(m,n)$, corresponding to the fast time scale of the Turing system. Then,
by Eq.~(\ref{EqEdot}), the system moves slowly along this nullcline according to
\begin{equation}
\partial_t E=\alpha_E-\frac{\kappa_{E}\,\alpha_B}{\kappa_B}\,f_{(m,n)}(L^2\,E)\,.
\end{equation}
Stability of steady-state patterns requires $\partial_E f_{(m,n)}>0$, which
can be shown to hold only for $E_{(m,n)}^{+}$, see
Fig.~\ref{SourceGradientScaling}(f).

Which branch $f_{(m,n)}$ is selected for arbitrary initial conditions by the fast Turing dynamics? 
This problem is formally equivalent to the stability of $(m,n)$-patterns in the Turing system without expander feedback as a function of system size $L$. 
From the analysis presented in Fig.~\ref{ModeStabPlot}(b),
we deduce that the (1,1)$^{+}$-pattern is the
only stable pattern in the blue region, 
which thus represents a basin of attraction.
Numerical analysis shows that the basin of attraction of the
(1,1)$^{+}$-pattern is even larger than the blue region and that this
pattern is stable also for nonadiabatic expander dynamics, see the trajectories in Fig.~\ref{SourceGradientScaling}(d).

In summary, the scaling mechanism for patterns and sources presented here
relies on expander molecules that dynamically adjust the 
degradation rates of morphogens in a Turing system.
Thereby, the expander controls the pattern length scales and the source size of the resulting Turing patterns.
The expander concentration is itself dynamic and is regulated by the concentrations of the Turing morphogens.
For the feedback introduced here, the relative source size at steady state is always independent of system size, 
see Eq.~(\ref{Eqrelsourceconst}).
We showed that a head-tail polarity pattern with a single source region 
scales as a function of system size, 
is stable with respect to perturbations, 
and regenerates in amputation fragments.

Regeneration of patterns after amputation can be understood as follows.
For a head fragment without a source, 
and hence no inhibitor production, 
the inhibitor level decreases, 
which decreases the expander degradation rate. 
Hence, the expander level increases.
For a tail fragment,
the inhibitor produced by the source spreads in a smaller system.
This implies higher inhibitor levels, which in turn decreases the source size. 
Only when the relative source size has fallen below its steady-state value, 
does the expander level increase. 
For head and tail fragments, the increasing expander level 
increases the degradation rate of activator and inhibitor, 
and thus scales down their pattern length scales.

For a given feedback scheme, the stability of fixed points depends on whether the source is fixed 
\cite{benzvi2010scaling,benzvi2011scaling,umulis2013mechanisms} or dynamic as in our case.
For example, two mutually suppressing concentration profiles (here: the inhibitor and the expander)
would not result in a stable pattern for a fixed source size, 
but yield a stable scaling pattern in our case, since the expander also effectively expands the source.

The minimal mechanism presented above allows for several generalizations.
First, the feedback of the Turing system on the expander level could be likewise implemented via the production rate, 
e.g., $\alpha_E\propto B$, instead of via the degradation rate $\beta_E=\kappa_E B$.
Then, scaling would require $\beta_A\propto1/E\,,\;\beta_B\propto1/E$, which yields analogous results.  
As a second possibility for pattern scaling, the feedback in Eq.~(\ref{EqbeE}) could also be mediated by $A$ instead of $B$, provided the expander diffuses sufficiently fast. More generally, similar results also follow for shuttling mechanisms for which $E$ adjusts both the degradation rates and diffusion coefficients of $A$ and $B$.  However, controlling only diffusion is not compatible with self-organized pattern scaling as presented here. Our mechanism relies on a size-dependent amplitude of morphogen profiles, which is lacking for pure diffusion control.

It is interesting to note that the flux $\beta_A A$ has a size-independent amplitude.
The spatial profile of this flux could provide a read-out of the scaling morphogen profiles 
independent of their amplitudes.\\ 

\textit{Conclusion.}
Motivated by biological examples of patterns that adjust to organism size 
\cite{gregor2005diffusion, umulis2013mechanisms,
wartlick2011dynamics, wartlick2011understanding,
benzvi2011expansion, hamaratoglu2011dpp,
benzvi2011scaling, newmark2002not},
we present a minimal, self-organized patterning system
that reliably establishes a head-tail pattern, scaled to match system size for a broad range of initial conditions. 
We extended a classical Turing system featuring local activation and lateral inhibition by a feedback loop, comprising a third diffusible molecule. 
The kinetics of this expander depends on the Turing patterns and feeds back on the Turing length scales.
Thereby, the expander effectively serves as a chemical size reporter. 
In contrast to earlier works on gradient scaling \cite{othmer1980scale, benzvi2010scaling,
wartlick2011dynamics, wartlick2011understanding,
benzvi2011expansion, benzvi2011scaling,
hunding1988size,ishihara2006turing}, this mechanism is fully self-organized. In particular, it does not rely on prepatterned sources or sinks.

In size-monitoring systems, as considered here, a key challenge relates to the simple fact that these obviously require long-range communication across the scale of the system. This implies a tradeoff between an upper size limit for scaling, and the time scale of pattern formation. 
Here, this time scale is set by morphogen diffusion and system size. 
For example, assuming a maximum diffusion coefficient of $100\, \mu$m$^2$/s and a maximum organism size of $20$ mm, relevant for the flatworms considered, we infer a patterning time scale of $3-30$ days, 
roughly consistent with the experimental range of $1-2$ weeks 
for the restoration of body plan proportions
after amputation \cite{newmark2002not,adell2010gradients}.
Note that transport processes such as active mixing could accelerate morphogen dispersal, and thus allow for faster pattern formation \cite{gregor2005diffusion}. In the minimal theory formulated here, no expander degradation occurs in the absence of the inhibitor. A basal degradation, independent of the inhibitor, would cap the expander concentration and thus set a lower size limit for scaling.

Our theory provides basic insight into principles of self-organized pattern scaling and accounts for key qualitative features of scalable patterning during flatworm regeneration and growth.
Three important signatures can be associated with the self-organized scaling mechanism introduced here:
(i) overall levels of morphogens depend on system size, 
(ii) morphogen degradation rates depend on system size, and 
(iii) the source size after amputation can exhibit a nonmonotonic dynamics.
These signatures provide explicit testable predictions 
regarding the regulatory dynamics of candidate patterning pathways such as Wnt signaling during regeneration and growth or degrowth in flatworms.
Interestingly, the expression of a Wnt activator (Wnt11-5) indeed displays a nonmonotonic dynamics during regeneration \cite{gurley2010expression}, reminiscent of signature (iii).
In the future, it will be important to quantify
spatial profiles of signaling molecules and degradation rates as a function of system size, 
which will allow us to test the generic concepts presented here.

\textit{Acknowledgments.}
We thank M.~K{\"u}cken, L.~Brusch, Y.~Quek, A.~Thommen, S.~Mansour and S.Y.~Liu for stimulating discussions.
S.W.~and B.M.F.~gratefully acknowledge support from the German Federal Ministry of Education and Research (BMBF), Grant No. 031 A 099.


\bibliography{/Users/admin/Documents/Publications/PaperTuringScaling,/Users/admin/Documents/Publications/PlanarianMorphogen,/Users/admin/Documents/Publications/DrosophilaMorphogen,/Users/admin/Documents/Publications/GradientScaling,/Users/admin/Documents/Publications/TuringSystems,/Users/admin/Documents/Publications/MorphogenTheory,/Users/admin/Documents/Publications/TuringMeinhardt,/Users/admin/Documents/Publications/TuringScaling,/Users/admin/Documents/Publications/PlanarianNeoblastsTurnover,/Users/admin/Documents/Publications/MiscOrganisms,/Users/admin/Documents/Publications/MiscPatternScaling,/Users/admin/Documents/Publications/PlanarianMisc}

\end{document}